41# The Preliminary design of DC Magnet Power Supply System for ITER Static Magnetic Field Test facility

Xi Deng, Ya Huang, Li Jiang, Ge Gao, Zhengyi Huang, Jie Zhang and Peng Wu*Abstract*—ITER (International Thermonuclear Experimental Reactor) static magnetic field (SMF) test facility requires a DC power supply with low voltage, high current, and high stability. Due to the limitation of switching loss, there is a contradiction between the output current capability and the output ripple. Large output current usually leads to low switching frequency, and low switching frequency will generate a large number of harmonics. To solve the problems, a topology based on the interleaving parallel buck converter is used and tested in this paper. Moreover, the topology is realized with only a small number of switching metal-oxide-semiconductor field effect transistors (MOSFETs). This article introduces the system design scheme and control method in detail. The analysis of harmonic and simulation are carried out. The validity of proposed scheme and control strategy were confirmed by experiments, the power supply system can supply large current of 15kA and has ability of low ripple.

Submit to: This work has been submitted to the IEEE for possible publication. Copyright may be transferred without notice, after which this version may no longer be accessible.## I. INTRODUCTION

The maximum static magnetic induction in the space around the ITER tokamak device exceeds 200mT. Such a high space magnetic field will affect the operational stability and reliability of the entire nuclear fusion device system. In order to study the impact of the ultra-high spatial magnetic field distribution of the ITER tokamak device on the device components during operation, high current, and high stability power supply is needed to establish a corresponding test magnetic field to test the key components of the tokamak device. The test facility mainly consists of inductance coil, cooling system, adjustable power supply, electronic device under test (EUT), and auxiliary equipment, as shown in Fig. 1 and Fig. 2.

The design of SMF magnet power supply faces challenge, due to the requirement of high current and low harmonics under high power density. To get high output current ability, the switching frequency of switching device is low, which leads to an increase in the output ripple, which in turn affects the stability of the power supply. Meantime, to maintain the stability of the system, the filter parameters of the power supply need to be increased, which reduce the power density of the system. For EAST (experimental advanced superconducting tokamak) in-vessel power supply, since the power supply requires four-quadrant operation, the H-bridge carrier phase shift (CPS) multilevel technology is applied to solve the contradiction between high power and low harmonic by increasing the equivalent switching frequency [1, 2]. But, the CPS multilevel H-bridge contains many switching tubes, so that its complexity has risen to a large extent.

This work was supported in part by the National Key Research & Development Plan 2017YFE0300401 and Comprehensive Research Facility for Fusion Technology (No. 2018-000052-73-01-001228). (Corresponding author: Li Jiang)
X. Deng are with the Institute of Plasma Physics, Chinese Academy of Sciences, Hefei 230031, China, and also with the University of Science and Technology of China, Hefei 230026, China (e-mail: xi.deng@ipp.ac.cn).
Y. Huang, L. Jiang, G. Gao, Z. Huang, J. Zhang, P. Wu are with the Institute of Plasma Physics, Hefei Institutes of Physical Science, Chinese Academy of Sciences, Hefei 230031, China (e-mail: ya.huang@ipp.ac.cn; jiangli@ipp.ac.cn; gg@ipp.ac.cn; hzy@ipp.ac.cn; zhangjie@ipp.ac.cn；wupeng@ipp.ac.cn).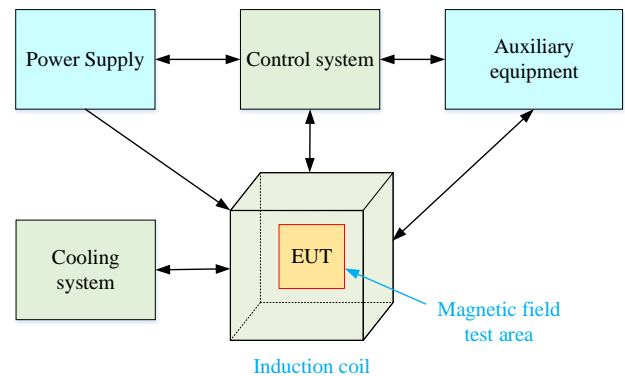

Fig. 1. Structure of ITER Static Magnetic Field Test facility

In this paper, a topology based on the interleaving parallel buck converter is proposed. This interleaving parallel buck converter can effectively achieve high current output and high stability at the same time by eliminate low-order harmonics and increase the equivalent switching frequency. The interleaving parallel buck topology is easy to achieve redundancy and ensure reliability. Moreover, this multilevel topology needs only a small number of switching tubes, which reduce the total cost of equipment. The validity of proposed scheme and control strategy were confirmed by simulation and experiments.

Fig. 2. Field equipment drawing of static magnetic field test equipment

I. THE TOPOLOGY OF SMF TEST FACILITY POWER SUPPLY

SMF test facility power supply has the characteristics of large output current and high output stability, which needs switching device with high switching frequency and high output current. IGCT (Integrated Gated Commutated Thyristor) and IEGT (Injection Enhanced Gate Transistor) can output large current and high voltage, but its switching frequency is low. High-voltage IGBT (Insulated Gate Bipolar Transistor) cannot output high current. Considering the output characteristics of the above switching devices, medium-voltage IGBT and MOSFET (Metal-Oxide-Semiconductor Field-Effect Transistor) are good choice. In this paper, MOSFET IXFK360N15T2 with 150A rated current and 10 kHz switching frequency is applied to ensure high resolution and high precision current control. Since the switching frequency of MOSFET is high, the dynamic current sharing problem of parallel MOSFET is difficult to solve. So, the topology of interleaving parallel buck converter, shown in Fig. 3, is adopted in this paper. Each buck circuit uses an independent industry standard rectifier module and is connected in parallel. The equivalent frequency exceeds 120 kHz. The interleaved trigger modulation method is described in details in the next section.

The power supply system of the SMF test facility consists of 24 power units. All 24 power units are connected in parallel to output 250A~13500A. Each power unit can convert the AC grid input into a DC constant current of a specified size. The output current of each power unit is 0~570A. The power unit is consisted of EMI filter, industry standard rectifier module, four-phase interleaving parallel buck modules, control and protection module, communication module, and auxiliary power system, as shown in Fig. 4. The main technical indexes of the system are shown in Table I.

TABLE I
MAIN TECHNICAL INDEXES OF THE SMF TEST FACILITY POWER SYSTEM

| Parameter | Value |
| --- | --- |
| Coil inductance $L_o$ (mH) | 4.5 |
| Coil resistance $R_o$ (m$\Omega$) | 2.3 |
| Input voltage $U_{ac}$ (V/Hz) | 380/50 |
| Nominal output current $I_{no}$ (kA) | 12.23 |
| Range of output current $I_o$ (kA) | 0.25~15 |
| Output current drift within 30 min | $\leq 0.5\%$ |
| Peak-peak current fluctuation | $\leq 0.5\%$ |
| Output current accuracy | $\leq 0.5\%$ |

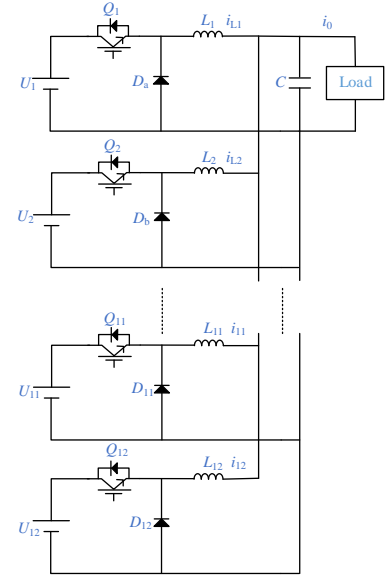

Fig. 3. Four-phase interleaving parallel buck converter topology

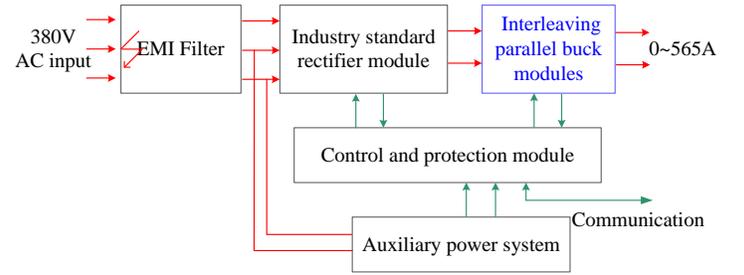

Fig. 4. Power supply Structure of ITER Static Magnetic Field Test facility

II. INTERLEAVING PARALLEL MODULATION METHOD

A. Modulation method based on interleaved trigger method

In the parallel power supply system, the switch device driving modes of each module are divided into three types, as shown in Fig. 5. Synchronous triggering refers to an identical driving signal for driving the respective switch modules. The total inductor ripple current is superimposed and increased in this case. Independent triggering method means that the switching frequency of each module is inconsistent. The total inductor ripple current is superimposed randomly and irregularly when using this method. As for interleaved trigger method, the switching frequency of each module is the same, and the drive signal has a regular phase difference. So, the current ripple of each module will cancel each other out. The harmonic of the parallel system will be reduced. The design adopts 12-phase interleaved trigger buck topology. For convenience, this paper takes 4-phase interleaved trigger buck topology as an example for subsequent analysis. The switching function of four-phase interleaving parallel modulation method is shown in Fig. 6. The phases of the driving signals of the switching device $Q_1$, $Q_2$, $Q_3$, $Q_4$ are shifted by 90°.

When the circuit operates, the output voltage of the circuit depends on the duty radio of the buck module. Since four buck module is connected in parallel, the output current is increases up to four times the single-module current amplitude. The equivalent switching frequency of the output current is four times that of the single-module. The total output current of the interleaving parallel buck modules exceeds 570A, and the equivalent frequency is 120 kHz. The circuit adjusts the output current by adjusting the output duty radio of the buck circuit.

Assuming that the inductance of each phase and the branch devices of the converter are ideal, the charge and discharge of the inductance are linear, take $0<DT\leq T/4$ working mode for analysis. In one cycle, the working process of the converter in continuous mode can be divided into eight parts, as shown in Fig.7.

During $t_0\sim t_1$, $T/4\sim t_2$, $T/2\sim t_3$, $3T/4\sim t_4$, the switch device of corresponding branches 1,2,3 and 4 are turned on, respectively, and the corresponding branch inductance is charged by the voltage source. The current flowing through this inductor increases. Inductance of other branches is discharged, and its current drops. The capacitor is charged.

At other times, all branch switch $Q_1$, $Q_2$, $Q_3$, $Q_4$ is closed, its branch inductance $L_1$, $L_2$, $L_3$, $L_4$ is discharged, and freewheeling is carried out through diode $D_a$, $D_b$, $D_c$, $D_d$. Capacitor is discharged.

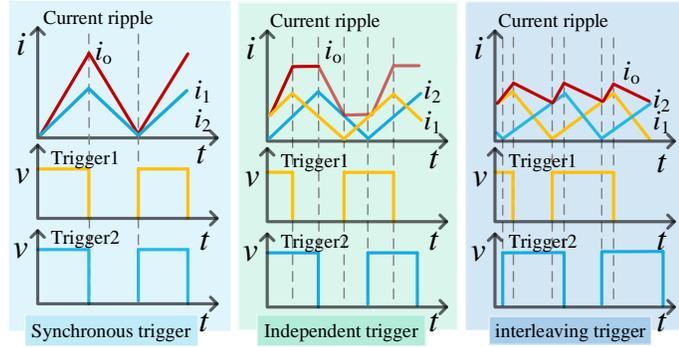

Fig. 5. Parallel operation mode

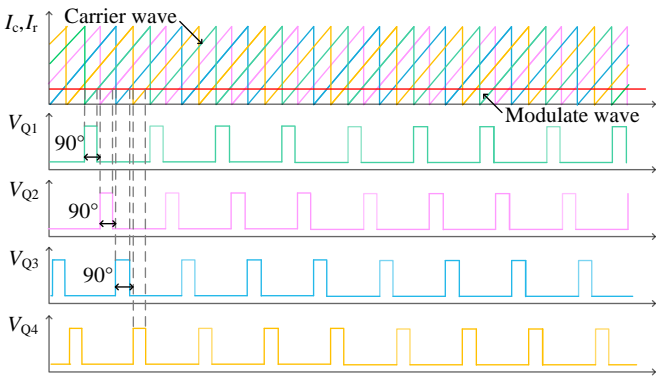

Fig. 6. Interleaving parallel modulation method·

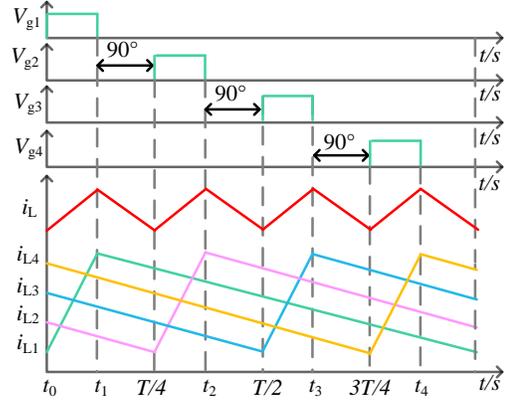

Fig. 7. Ideal inductor current and drive signal

### B. Analysis of inductor current ripple and harmonic

According to the above analysis, when $0<DT\leq T/4$, the working process of the circuit is repetitive, with $T/4$ as a cycle. When the circuit is analyzed in the first $T/4$ period, it can be seen that the switch $Q_1$ is turned on during $t_0\sim t_1$, and the currents of the inductors $L_2$, $L_3$, and $L_4$ are all reduced; during $t_1\sim T/4$, all the switches are turn off. All the inductor currents are reduced. Assuming that each phase is working in continuous mode, the duty cycle of all phases, the delay of the drive signal, the turn-on voltage drops of the power device, and the parasitic resistance of each phase are the same. It can be deduced that the inductor current ripple is of each path is:

$$\begin{cases} \Delta i_{L1} = \dfrac{V_{in}-V_O}{L_1}DT = \dfrac{V_{in}(1-D)D}{L}T \\ \Delta i_{L2} = \Delta i_{L3} = \Delta i_{L4} = -\dfrac{V_O}{L}DT = -\dfrac{V_{in}D^2}{L}T \end{cases} \quad (1)$$

Where $\Delta i_{L1}$, $\Delta i_{L2}$, $\Delta i_{L3}$, $\Delta i_{L4}$ represents the inductor current ripple of each phase. After linear superposition, the total output current ripple of the four-phase interleaved parallel buck chopper circuit can be obtained as:

$$\Delta i_{L1} = \dfrac{V_{in}}{Lf}(1-4D)D \quad (2)$$

When the duty cycle in other cases, shown in Fig. 8, the corresponding current ripple is obtained as follows.

$$\begin{cases} \Delta i_{L2} = \dfrac{V_{in}}{Lf}(2-4D)(D-\dfrac{1}{4}) & (\dfrac{1}{4}<D\leq \dfrac{2}{4}) \\ \Delta i_{L3} = \dfrac{V_{in}}{Lf}(3-4D)(D-\dfrac{1}{2}) & (\dfrac{2}{4}<D\leq \dfrac{3}{4}) \\ \Delta i_{L4} = \dfrac{V_{in}}{Lf}(4-4D)(D-\dfrac{3}{4}) & (\dfrac{3}{4}<D\leq 1) \end{cases} \quad (3)$$

The total inductor current ripple peak value of $N$-phase interleaved trigger buck topology is:

$$\Delta I_{sum}(D,N) = \dfrac{V_{in}}{Lf}(D-\dfrac{m}{N})(1+m-N\cdot D) = \dfrac{V_{in}}{Lf}\cdot K \quad (4)$$

where $m$ is *floor* ($D\cdot N$), $K$ is the function of the normalization coefficient of the output current ripple.

Fig. 9 shows the function diagram of coefficient $K$, which is a function of the number of parallel phases $N$ and duty cycle $D$.

The characteristics of the output current ripple can be qualitatively estimated from the picture. By using interleaved parallel structure, the current ripple presents a symmetrical distribution and the total output current ripple is significantly reduced. The output current ripple is zero at some points, and the ripple increases with the increase of the number of parallel phases. Under the conditions of open-loop ideal parameters, such as four phases, the current ripple is zero when the duty cycle is 0.25, 0.5 and 0.75. In practical closed-loop applications, the regulator automatically adjusts duty ratio to adapt to disturbances of load or power, so it does not meet the working conditions of zero ripple. At the same time, near the point where the current ripple is zero, the current ripple amplitude corresponding to different phase numbers changes very little. Fig. 10 shows the ratio of current ripple of different phases interleaving buck converter to current ripple of single buck converter. When the duty cycle is pretty small or pretty large, the ripple of the output current can be significantly reduced by increasing the number of phases of the converter.

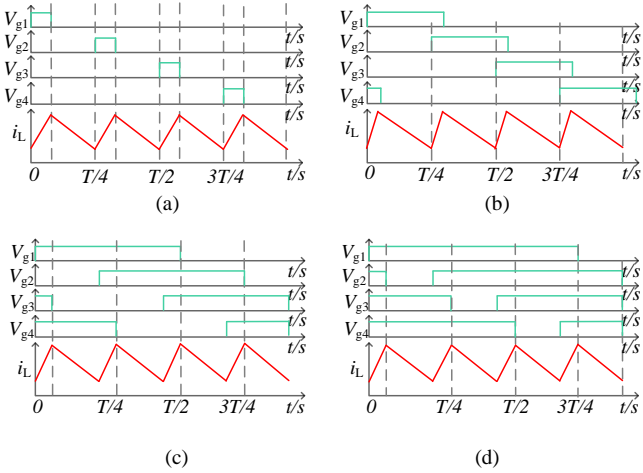

Fig. 8. Inductor current and drive signal(a) 0<DT≤T/4, (b) T/4<DT≤T/2, (c) T/2<DT≤3T/4, (d)3T/4<DT≤T.

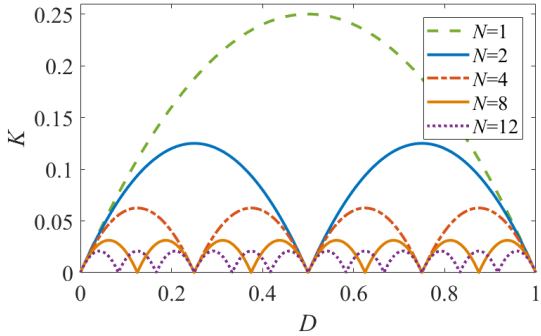

Fig. 9. Normalization coefficient of the output current ripple under different phase.

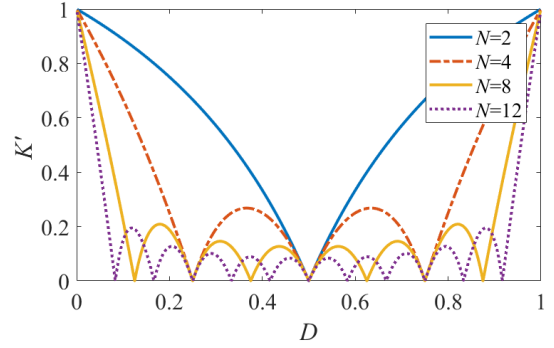

Fig. 10.The ratio of total output current ripple of different phases interleaving buck converter to current ripple of single buck converter

### C. Fourier analysis of Current harmonic of interleaving parallel buck topology

The inductor current ripple waveform of single buck converter in continuous conduction mode (CCM) is shown in Fig. 11.The time-domain representation for one period of the inductor current waveform is given by

$$i(t)' = \begin{cases} \dfrac{\Delta i}{DT} \cdot t + I_{out} - \dfrac{\Delta i}{2} & 0 \leq t \leq DT \\ -\dfrac{\Delta i}{(1-D)T} \cdot t + I_{out} + \dfrac{\Delta i}{2} & DT \leq t \leq T \end{cases} \quad (5)$$

where $I_{out}$ is the mean value of the inductor current. $\Delta i$ is the fluctuation value of the inductor current.

In order to facilitate the calculation of Fourier series, the coordinate axis in

Fig. 11 is shifted to make the inductor waveform symmetric about origin. The graph after coordinate transformation is shown in the Fig. 12[3]. The waveform amplitude is between -$\Delta i$ and $\Delta i$. The intersection of the waveform and the horizontal axis is 0, $T/2$, $T$. The time-domain representation for one period of inductor current waveform after coordinate shift is given by:

$$i(t) = \begin{cases} \dfrac{\Delta i}{DT} \cdot t, & 0 \leq t \leq \dfrac{DT}{2} \\ \dfrac{\Delta i}{2} - \dfrac{\Delta i}{(1-D)T/2} \cdot \left(t - \dfrac{DT}{2}\right), & \dfrac{DT}{2} \leq t \leq T - \dfrac{DT}{2} \\ \dfrac{\Delta i}{DT} \cdot (t-T), & T - \dfrac{DT}{2} \leq t \leq T \end{cases} \quad (6)$$

The Fourier series of the above inductor current is given as:

$$i_n(t) = \dfrac{a_0}{2} + \sum_{k=1}^{\infty}[a_k \cdot \cos(k(\omega t - \phi_n)) + \ldots + b_k \cdot \sin(k(\omega t - \phi_n))] \quad (7)$$

where $n$ refers to the $n$th buck circuit and $k$ is the harmonic order. $\phi_n$ denotes the phase shift angle of the $n$th buck circuit. The form is related by

$$\phi_n = (n-1) \cdot \dfrac{2\pi}{N} \quad (8)$$

Since the function *i(t)* is an odd function, the Fourier series of an odd function will not contain a cosine term, but only a sine term. So,

$$a_0 = 0 \quad (9)$$
$$a_k = 0 \quad (10)$$

The coefficient $b_k$ is as follows:

$$b_k = \frac{2}{T}\int_0^T i(t)\sin(k\omega t)dt = -\frac{\Delta i(-1)^k}{k^2 D(1-D)\pi^2}\sin[k(1-D)\pi] \quad (11)$$

The Fourier series (7) is converted into cosine standard form as follows:

$$i(t) = \frac{a_0}{2} + \sum_{k=1}^{\infty} A_k \cdot \cos(k(\omega t - \phi_n) - \varphi_k) \quad (12)$$

When applying the substitution

$$\frac{a_0}{2} = 0 \quad (13)$$

$$A_k = \sqrt{a_k^2 + b_k^2} = b_k \quad (14)$$

$$\varphi_k = \operatorname{atan2}(b_k, a_k) = \pm\frac{\pi}{2} \quad (15)$$

$$\theta_{nk} = k\phi_n + \varphi_k = k\phi_n \pm \frac{\pi}{2} \quad (16)$$

We can get equation (17) as follows:

$$i(t) = \sum_{k=1}^{\infty} b_k \cdot \cos(k\omega t - k\phi_n \pm \frac{\pi}{2})$$
$$= \sum_{k=1}^{\infty} b_k \cdot \cos(k\omega t - \theta_{nk}) \quad (17)$$

In order to simplify the analysis, equation (17) can be expressed as a phasor

$$A_k \cdot e^{-j\theta_{nk}} = (b_k \cdot e^{-j\varphi_k}) \cdot e^{-jk\phi_n} \quad (16)$$

The topology is four-phase buck converter, *N*=4. The phasor in (16) can be deduced as follows:

①When *k*=1, the phasor of the four-phase buck converter is

$(b_1 \cdot e^{-j\varphi_1}) \cdot e^{-j\frac{\pi}{2}}$, $(b_1 \cdot e^{-j\varphi_1}) \cdot e^{-j\pi}$, $(b_1 \cdot e^{-j\varphi_1}) \cdot e^{-j\frac{3\pi}{2}}$, $(b_1 \cdot e^{-j\varphi_1}) \cdot e^{-j2\pi}$

②When *k*=2, the phasor of the four-phase buck converter is

$(b_2 \cdot e^{-j\varphi_2}) \cdot e^{-j\pi}$, $(b_2 \cdot e^{-j\varphi_2}) \cdot e^{-j2\pi}$, $(b_2 \cdot e^{-j\varphi_2}) \cdot e^{-j3\pi}$, $(b_2 \cdot e^{-j\varphi_2}) \cdot e^{-j4\pi}$

③When *k*=3, the phasor of the four-phase buck converter is

$(b_3 \cdot e^{-j\varphi_3}) \cdot e^{-j\frac{3\pi}{2}}$, $(b_3 \cdot e^{-j\varphi_3}) \cdot e^{-j3\pi}$, $(b_3 \cdot e^{-j\varphi_3}) \cdot e^{-j\frac{9\pi}{2}}$, $(b_3 \cdot e^{-j\varphi_3}) \cdot e^{-j6\pi}$

④When *k*=4, the phasor of the four-phase buck converter is

$(b_4 \cdot e^{-j\varphi_4}) \cdot e^{-j2\pi}$, $(b_4 \cdot e^{-j\varphi_4}) \cdot e^{-j4\pi}$, $(b_4 \cdot e^{-j\varphi_4}) \cdot e^{-j6\pi}$, $(b_4 \cdot e^{-j\varphi_4}) \cdot e^{-j8\pi}$

The relative position and amplitude of inductor current harmonic of four-phase buck converter is shown in Fig. 13. The 1th, 2th, 3th, and 4th harmonics are interleaved to cancel each other, while the 4th harmonics are superimposed on each other. Similarly, the 4*N*th, (4*N*+1)th, (4*N*+2)th, (4*N*+3)th are staggered and offset, while the 4*N*th harmonics are superimposed.

Therefore, there is only the 4*N*th harmonic in the total input current. The fundamental harmonic of the total input current starts from the 4th harmonic, which reduces the amplitude of the fundamental harmonic, improves the frequency of the fundamental harmonic, and improves the EMI characteristics of the system.

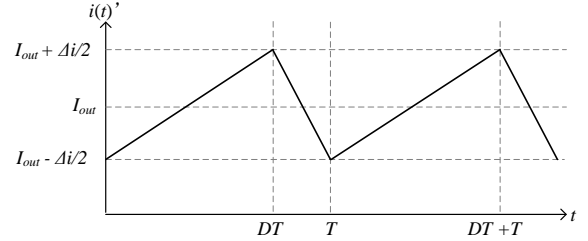

Fig. 11. Inductor current waveform for a buck converter in CCM

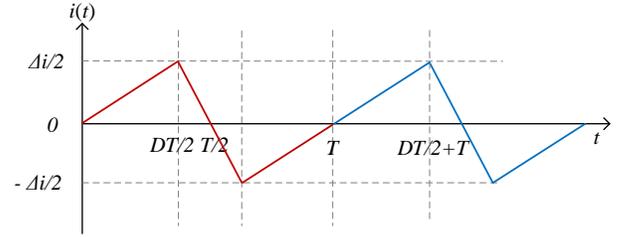

Fig. 12. Inductor current waveform for a buck converter in CCM after time-shifted

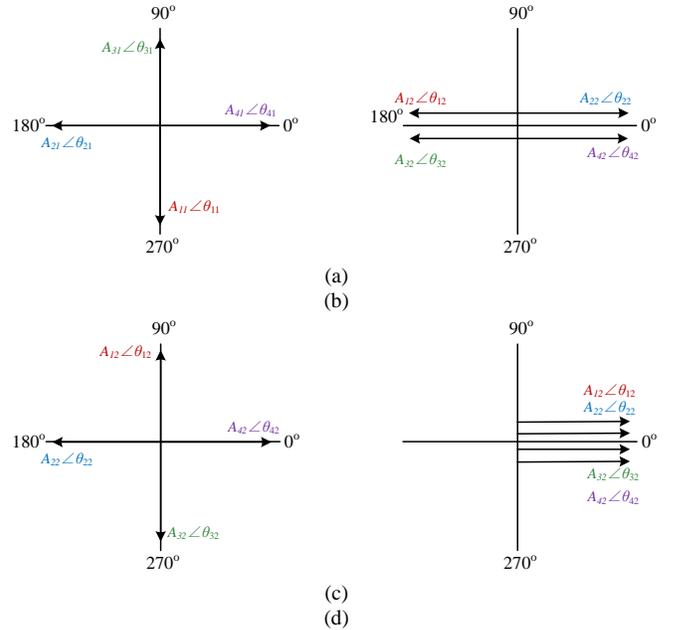

Fig. 13. The relative position and amplitude of each phase inductor current harmonic of four-phase buck converter (a) k=1; (a) k=2;(c) k=3;(d) k=4.

## II. TEST RESULTS

To verify the correctness of the calculation results, a test system platform is built, as shown in Fig. 14.

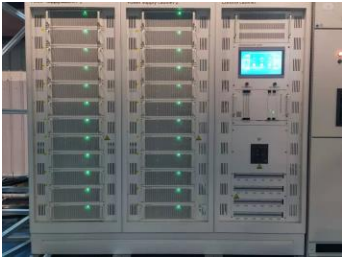

Fig. 14. Power supply cabinet of ITER SMF test facility.

## III. Conclusion

To ensure the thermal stability of the magnetic field immunity test system, the electromagnetic heat flux coupling analysis is needed.

## Acknowledgment

This work was supported by the International Thermonuclear Experimental Reactor (ITER) Organization.